\documentclass[useAMS,usenatbib,intlimits,centertags,usegraphicx]{mn2e}
\usepackage{amsmath,amssymb}
\usepackage[latin1]{inputenc}

\newcommand{\beq}{\begin{equation}}
\newcommand{\eeq}{\end{equation}}

\begin{document}

\title{Uniformly Rotating Rings in General Relativity}

\author[Fischer, Horatschek \& Ansorg]
 {Thomas Fischer$^1$,
  Stefan Horatschek$^1$ 
  and
  Marcus Ansorg$^2$\\
  $^1$Theoretisch-Physikalisches Institut, University of Jena,
                Max-Wien-Platz 1, 07743 Jena, Germany\\ 
  $^2$Max-Planck-Institut f\"ur Gra\-vi\-ta\-ti\-ons\-phy\-sik, 
                Albert-Einstein-Institut, Am M\"uhlenberg 1, 14476 Golm, Germany}

\pagerange{\pageref{firstpage}--\pageref{lastpage}} \pubyear{2005}

\maketitle

\label{firstpage}

\begin{abstract}
	In this paper, we discuss general relativistic, self-gravitating and uniformly 
	rotating perfect fluid bodies with a toroidal topology (without central object).
	For the equations of state describing the fluid matter we consider polytropic 
	as well as completely degenerate, perfect Fermi gas models.
	We find that the corresponding configurations possess similar properties to the 
	homogeneous relativistic Dyson rings. On the one hand, there exists no limit to the 
	mass for a given maximal mass-density inside the body. 
	On the other hand, each model permits a quasistationary transition to the 
	extreme Kerr black hole.
\end{abstract}

\begin{keywords} relativity -- gravitation -- methods: numerical -- stars: rotation. \\

preprint number: AEI-2005-112
\end{keywords}

\section{Introduction}
	The first studies of uniformly rotating, axisymmetric and self-gravitating, 
	toroidal perfect fluids in equilibrium date
	back to the 19th century, when 
	\citet{Kowalewsky85}, \citet{Poincare85b,Poincare85c,Poincare85d} and 
	\citet{Dyson92,Dyson93} considered thin homogeneous rings in 
	Newtonian gravity. The analytical expansion found by Dyson is a fourth order
	approximation of the solution to the corresponding free boundary value problem. 
	Numerical investigations of these Dyson rings have been carried out by 
	\citet{Wong74}, \citet{ES81} and \citet*{AKM03}. 
	In their studies, Eriguchi \& Sugimoto were able to confirm 
	a conjecture by \citet{Bardeen71} about a continuous parameter transition 
	of the Dyson ring sequence to the well known Maclaurin spheroids. 
	The analogues of these rings in Einsteinian gravity 
	(the ``relativistic Dyson rings'')  were treated by \citet*{AKM03c}.
	
	\citet{Hachisu86} extensively studied both spheroidal and toroidal 
	self-gravitating Newtonian fluid bodies (uniformly as well as differentially rotating) 
	governed by different equations of state, among them polytropes 
	(with polytropic index $N$ varying from 0 to 4) and white dwarves (described by 
	Chandrasekhar's equation of a relativistic, completely degenerate, perfect Fermi gas).
	In particular, he presented the critical configurations which rotate at the mass-shedding 
	limit and mark the endpoint of specific sequences. These investigations showed that, 
	in contrast to the uniformly rotating homogeneous models, for all other equations of state 
	considered, there exist disjoint Newtonian sequences of spheroidal and toroidal shape. 

	Proceeding to the relativistic realm, the uniformly rotating homogeneous bodies were also found 
	to form disjoint classes\footnote{They are only connected through specific bifurcation points 
	of the Maclaurin sequence at which this sequence becomes secularly unstable with respect to axisymmetric 
	perturbations and the post-Newtonian expansion fails.}. The corresponding complete picture in 
	Newtonian and Einsteinian gravity is presented in \citet{AKM03} and 
	\citet{AFKMPS04}.

	It is the aim of this paper to extend Hachisu's work to Einsteinian gravity. We study 
	uniformly rotating polytropes (with polytropic indices $N$ varying from 0 to 4)
	as well as neutron gas configurations with a toroidal shape. The neutron gas 
	bodies are motivated by Chandrasekhar's equation of state applied to a neutron star.  
	Like Hachisu, we identify the mass-shedding configurations that bound our solution space. 

	Uniform rotation is widely used to describe rotating neutron stars. 
	In particular, when restricting oneself to stationary solutions, it is all but
	necessary to consider this simple rotation law,
	since any realistic matter would gradually smooth out any imposed 
	differential rotation by means of some dissipative processes.
	In addition, note that through differential rotation a free function describing the angular 
	velocity as it depends on radius would enter into the game and a concise representation 
	of some basic properties of the rings in question (as intended in our paper)
	would be difficult. Therefore, in this paper we focus our attention on uniform rotation.
	
	As proved by \citet{Meinel04}, 
	only the extreme Kerr black hole can be the black hole limit of rotating perfect fluid bodies
	in equilibrium. In \citet{AKM03c} a numerical example of such a transition to the extreme Kerr 
	black hole was found for the relativistic Dyson rings.
	Here, we follow this line and find that such a transition is a general feature of 
	ring configurations, i.e.~it was found for all models studied.   

	Our numerical investigations are based on the highly accurate multi-domain,
	pseudo-spectral method described in \citet*{AKM03b}, which allows us to calculate even
	the limiting configurations very precisely. 

	In the units used, the speed of light as well as Newton's
	constant of gravity are equal to~1.
\section{Metric Tensor and Field Equations} \label{Metric}
	The line element for a stationary, axisymmetric spacetime describing the gravitational field of 
	a uniformly rotating perfect fluid body can be written in Lewis-Papapetrou coordinates as follows:
	\beq
		ds^2=e^{2\alpha}(d\varrho^2+d\zeta^2)+W^2e^{-2\nu}(d\varphi-\omega\,dt)^2-e^{2\nu}dt^2\,.
	\eeq
	We define these coordinates $(\varrho,\zeta,\varphi,t)$ uniquely by the requirement that all 
	metric functions and their first derivatives be continuous everywhere, in particular at the 
	fluid's surface. 

	For a perfect fluid body rotating with the uniform angular velocity $\Omega$, 
	the following boundary condition at the fluid's surface holds: 
	\beq
		e^{2\nu}-W^2(\omega-\Omega)^2e^{-2\nu}=\text{constant}=(1+Z_0)^{-2}.
	\eeq
	The constant $Z_0$ is the relative redshift measured at infinity of photons 
	emitted from the body's surface that do not carry angular momentum\footnote{No
	angular momentum means $\eta_ip^i=0$ ($p^i$:
	four-momentum of the photon, $\eta^i$: Killing vector corresponding to axisymmetry).}.

	Apart from the above boundary condition, regularity of the metric along the axis 
	$\varrho=0$ is required as well as reflectional symmetry with respect to the equatorial 
	plane $\zeta=0$. Taking the interior and exterior field equations, the transition conditions at the 
	fluid's surface, the above regularity requirements as well as the asymptotic behaviour 
	at infinity, we obtain a complete free-boundary problem to be solved. In particular, the unknown
	toroidal shape of the fluid body enters into this set of equations. 

	Note that for isolated general relativistic fluid bodies, the regularity condition at 
	infinity is usually asymptotic flatness. However, the situation becomes more complex 
	as the transition to the extreme Kerr black hole is considered, see \citet{Meinel02}. 
	In this limiting process the ring shrinks down to the coordinate origin and the resulting 
	exterior geometry is given by the extreme Kerr metric outside
	the horizon. On the contrary, a different non-asymptotically flat limit is encountered if
	the coordinates are rescaled in such a way that the extension of the ring remains finite. 
	In this case, the behaviour of the gravitational potentials at infinity
	is determined by the ``extreme Kerr throat geometry'' \citep{BH99}. 
\section{The Equations of State}
	In order to consider a specific model of a rotating fluid configuration it is essential to 
	describe the matter inside the body by a characteristic equation of state. In this paper we 
	treat polytropic as well as completely degenerate, perfect Fermi gas models. The 
	corresponding equations of state relate the mass-energy density $\mu$ and the pressure $p$,
	which enter the field equations through the energy-momentum tensor. It reads for a perfect fluid
	as follows:
	\beq
		T_{ik}=(\mu+p)u_iu_k+pg_{ik},
	\eeq
 	with $u^k$ being the four-velocity of the fluid elements. 
	
	For so-called isentropic models, especially in the zero temperature limit, a 
	rest-mass density $\mu_{\text B}$ can be derived from
	\beq
		\label{mu_B}
		\frac{d\mu_{\text B}}{\mu_{\text B}}=\frac{d\mu}{\mu+p}\,.
	\eeq
	The physical quantities $\mu,\mu_{\text B}$ and $p$ allow us to define
	a gravitational mass $M$ as well as a rest mass $M_0$, which we will use 
	as characteristic parameters to describe a specific fluid configuration.
	Note that the constant of integration in (\ref{mu_B}) is fixed by the requirement that 
	\beq
		\mu/\mu_{\text B}\to 1 \quad\text{as}\quad p\to 0.
	\eeq
	\subsection{Polytropic Models}
		The relativistic polytropic equation of state was given by
		\citet{Tooper65}:
		\beq
			\mu=Np+(p/K)^{N/(N+1)}.
		\eeq
		Here $N$ and $K$ are the polytropic index and polytropic constant respectively. 
		In this paper we vary $N$ within the interval $[0,4]$. 
		For small values, the equation of state is called ``stiff'', corresponding to a 
		strong fall-off of the mass densities in the vicinity of the fluid's surface.
		In particular, $N=0$ describes the homogeneous model in which the mass densities 
		($\mu$, $\mu_{\text B}$) jump at the surface. On the other hand, we obtain 
		a ``soft'' equation of state for 
		larger $N$, resulting in flat density profiles close to the fluid's boundary.

		Note that we may use the fundamental length $K^{N/2}$ in order to introduce dimensionless 
		quantities, 
		\beq
			\bar M=K^{-N/2}M,\quad\bar\Omega=K^{N/2}\Omega,
		\eeq
		see \citet{NSGE98}. 
	\subsection{Chandrasekhar's Equation of State}
		\begin{figure*} 
		\begin{center}
			\includegraphics[scale=0.8]{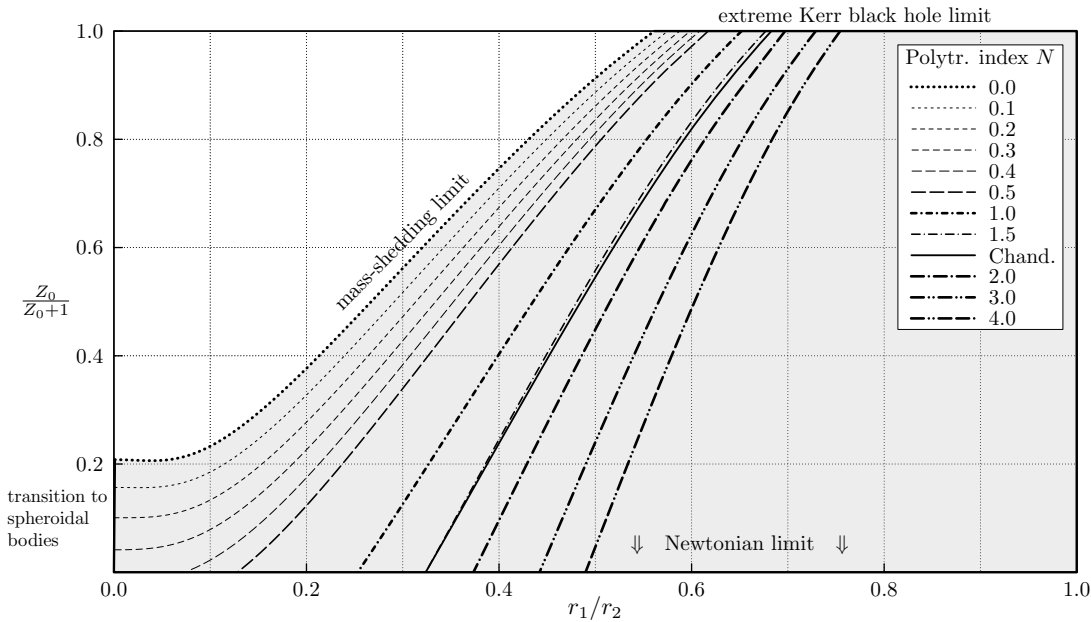}
		\end{center}
		\vspace{-1\baselineskip}
		\caption{\label{Abb1}	
		Parameter space of the rings considerd in the $r_1/r_2$-$Z_0/(Z_0+1)$ plane.
		Each parameter region corresponding to a particular equation of state
		is bounded from the left by a mass-shedding sequence. Note that the 
		transition to the extreme Kerr black 
		hole is exhibited for all matter models studied.
		}	
		\end{figure*}
	
	        \label{ChandEOS}
		Chandrasekhar's equation of state describes a relativistic, completely degenerate, 
		perfect Fermi gas \citep{OV39}. It is given implicitly by
		\newcommand\sqrtx[1]{\sqrt{\smash{#1}\rule{0mm}{1.8ex}}}
		\begin{align}
			\mu &= K_{\text C}\left[(6x_{\text F}^3+3x_{\text F})\sqrtx{1+x_{\text F}^2} 
						-3\ln\big(x_{\text F}+\sqrtx{1+x_{\text F}^2}\,\big)\right],\\
			p   &= K_{\text C}\left[(2x_{\text F}^3-3x_{\text F})\sqrtx{1+x_{\text F}^2} 
						+3\ln\big(x_{\text F}+\sqrtx{1+x_{\text F}^2}\,\big)\right].
		\end{align}
		Here the constant $K_{\text C}$, 
		\beq
			K_{\text C}=\frac{m^4(2s+1)}{48\pi^2\hbar^3}, 
		\eeq
		is derived from the mass $m$ and the spin $s$ of the fermions. Therefore, taking a neutron gas, 
		the equation of state is completely fixed and does not contain any free parameters.

		
		Note that $x_{\text F}=p_{\text F}/m$ is the dimensionless Fermi momentum of the gas. In the
  		non-relativistic limit, $x_{\text F}\ll 1$, the equation of state approaches the polytropic 
		equation of state with $N=3/2$.

\section{Results}
	The studies of \citet{Hachisu86} provide an extensive overview of 
	Newtonian ring configurations. For a number of equations of state the characteristic 
	features of the corresponding ring sequences are illustrated by means of specific 
	parameter diagrams and tables. Moreover, the shapes of representative examples
	are displayed in cross-section.
	
	With the transition to Einsteinian gravity we have an additional
	relativistic parameter at our disposal which makes it difficult to give a similar detailed
	overview over the corresponding relativistic picture. We focus therefore our attention
	on the important question of parametric transitions of the relativistic 
	objects to characteristic limiting configurations. As in \citet{AKM03c} we identify 
	the mass-shedding as well as the extreme Kerr black hole limit for each equation of state 
	considered. For the polytropes with $N=1$, we provide three parameter diagrams displaying the 
	characteristic behaviour of representative physical quantities of the corresponding fluid
	bodies. Finally, we consider an exemplary sequence of Fermi gas fluid bodies with prescribed
	rest mass $M_0=24M_\odot$ that terminates at the extreme Kerr black hole limit. 
	\subsection{Parameter Space of Ring Configurations}

	We may characterise each fluid body governed by a prescribed equation of state 
	by two physical parameters. In figure \ref{Abb1}, the relativistic quantity 
	$Z_0/(1+Z_0)$ is plotted against the ratio $r_1/r_2$ of the inner to outer (coordinate) radius
	of the rings. Note that $Z_0/(1+Z_0)$ vanishes in 
	the Newtonian limit and tends to unity in the extreme Kerr black hole limit.

	The relativistic Dyson rings (described by the polytropic constant $N=0$)
	possess parameter pairs that are located in the grey shaded area. 
	Apart from the limit of infinitely thin rings ($r_1/r_2\to 1$), this area is bounded by 
	(1) the Dyson rings in Newtonian gravity, 
	(2) configurations at the transition to spheroidal topology,
	(3) critical configurations that rotate at the mass-shedding limit\footnote{A mass-shedding limit is given 
	if a fluid particle at the surface of the body moves with the same angular velocity as a test particle at that 
	spatial point. The corresponding geometrical shape of the fluid body possesses 
	a cusp there.} and 
	(4) ultrarelativistic configurations at the transition to the extreme Kerr black hole 
	(see section \ref{KerrBH}).

	If we increase the polytropic index $N$, we find that the mass-shedding sequence moves 
	sideways. However, the corresponding areas are still bounded by the same types of limiting curves, among them
	the Newtonian sequences that were studied by Hachisu. Note that from a critical
	 $N_{\text{crit}}\approx 0.35$ on, there is no continuous transition to spheriodal bodies, as exhibited by the fact that the mass-shedding sequence 
	terminates at the Newtonian limit.

	In figure \ref{Abb1}, the parameter region with respect to Chandrasekhar's equation of state is also shown.
	Since these configurations have the same Newtonian limit as the polytropic rings with $N=3/2$, 
	the two corresponding mass-shedding sequences meet at the abscissa. We see that the overall deviation of these
	two curves remains small throughout the entire relativistic regime.
	\subsection{Polytropic Rings with $\bmath{N=1}$}

	\begin{figure} 
	  \begin{center}
	    \includegraphics[scale=0.8]{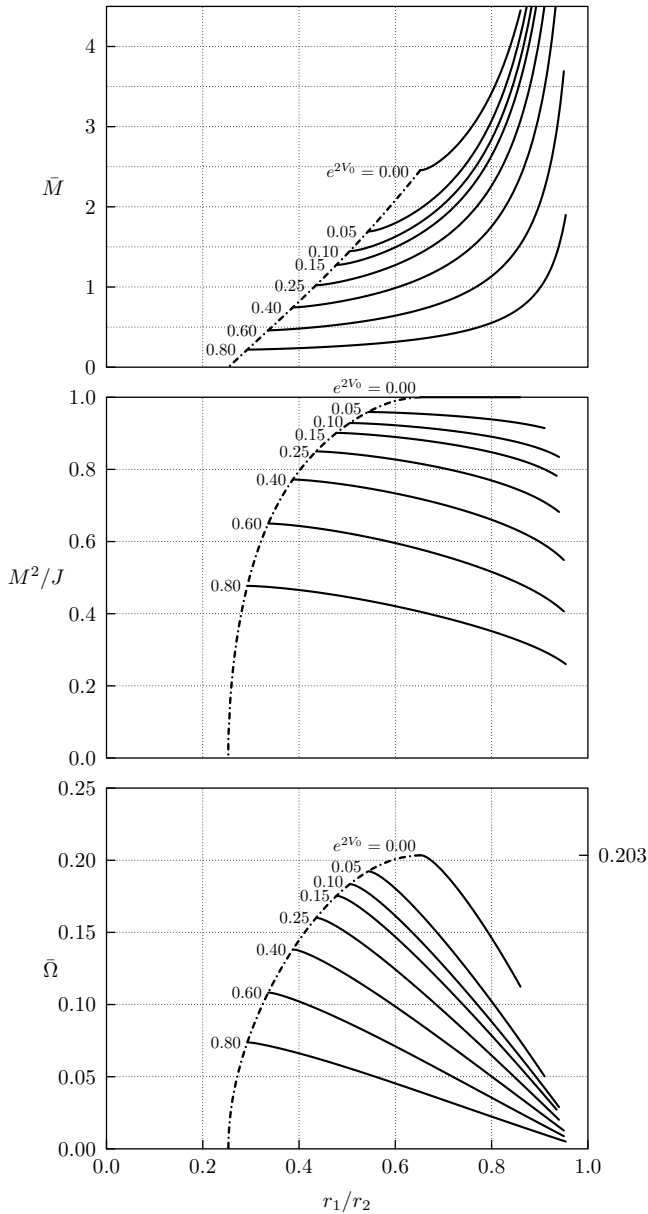}
	  \end{center}
	  \vspace{-1\baselineskip}
	  \caption{\label{Abb2}
	           For polytropic rings with $N=1$, 
		   the dimensionless gravitational mass $\bar M$, 
	           the (dimensionless) quantity $M^2/J$ and
		   the dimensionless angular velocity $\bar\Omega$ 
		   are plotted against the radius ratio $r_1/r_2$ for 
		   certain values of $e^{2V_0}=(1+Z_0)^{-2}$. 
		   The dashed lines describes the mass-shedding sequence.
			}
	\end{figure}
	
	As illustrated in figure \ref{Abb1}, for the different matter models very similar results emerge. 
	Let us, as an example, consider the class of relativistic polytropic rings with the polytropic 
	index $N=1$ in more detail. 
	
	Figure \ref{Abb2} shows the quantities $\bar M$, $M^2/J$ ($J$: angular momentum) and $\bar\Omega$ as 
	functions of the radius ratio. The solid lines represent sequences of configurations with a 
	constant prescribed value for the redshift ($Z_0=e^{-V_0}-1$). They terminate at the dashed line, 
	which describes the mass-shedding sequence (cf. figure \ref{Abb1}).
	
	As with the relativistic Dyson rings, we find that the (dimensionless) mass $\bar M$ grows to infinity 
	in the thin ring limit. 
	This result is also valid for the Dyson rings in Newtonian gravity, which can be proved analytically.
	In  particular, for fixed $V_0$ we find that
	\beq
	 \lim _{\sigma\to 0}\left[\bar M\sigma (-\ln\sigma)^{3/2}\right]=\sqrt{-\frac{32\pi}{125}V_0^3}
	\eeq
	 where $\sigma=(r_2-r_1)/(r_2+r_1)$.
	
	This subtle limit seems to possess unique features independent of the specific 
	equation of state  being considered. 
	Note that we find the same typical behaviour of $\bar M$ for 
	both stiff and soft equations of state.
	We plan a more thorough investigation of this issue in a subsequent 
	publication.
	\subsection{Quasistationary Transition to the Extreme Kerr Black Hole}\label{KerrBH}

	\begin{figure} 
	  \begin{center}
	    \includegraphics[scale=0.8]{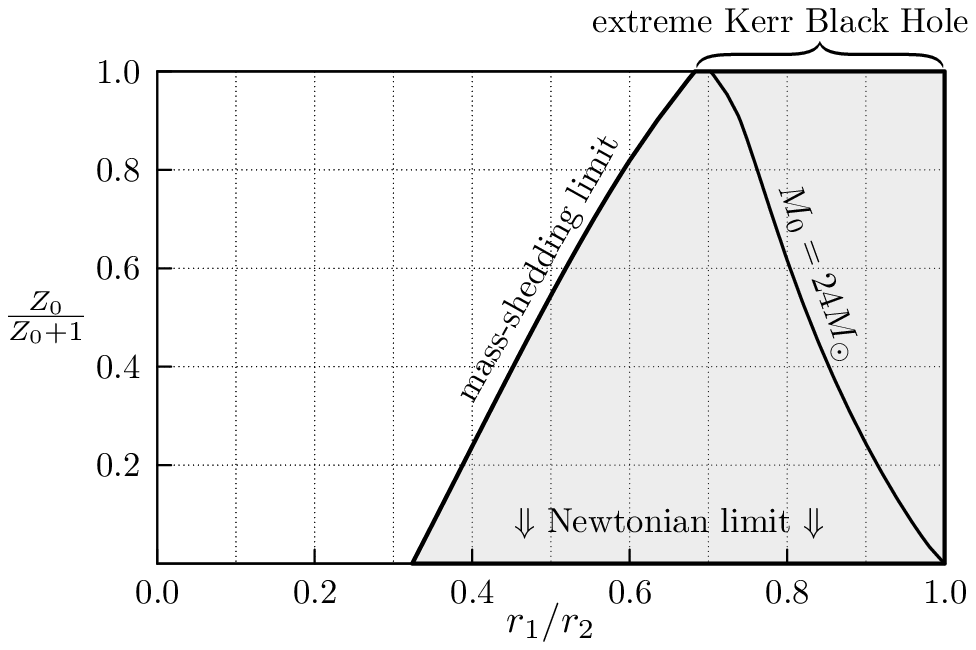}
	    \par\medskip
	    \includegraphics[scale=0.8]{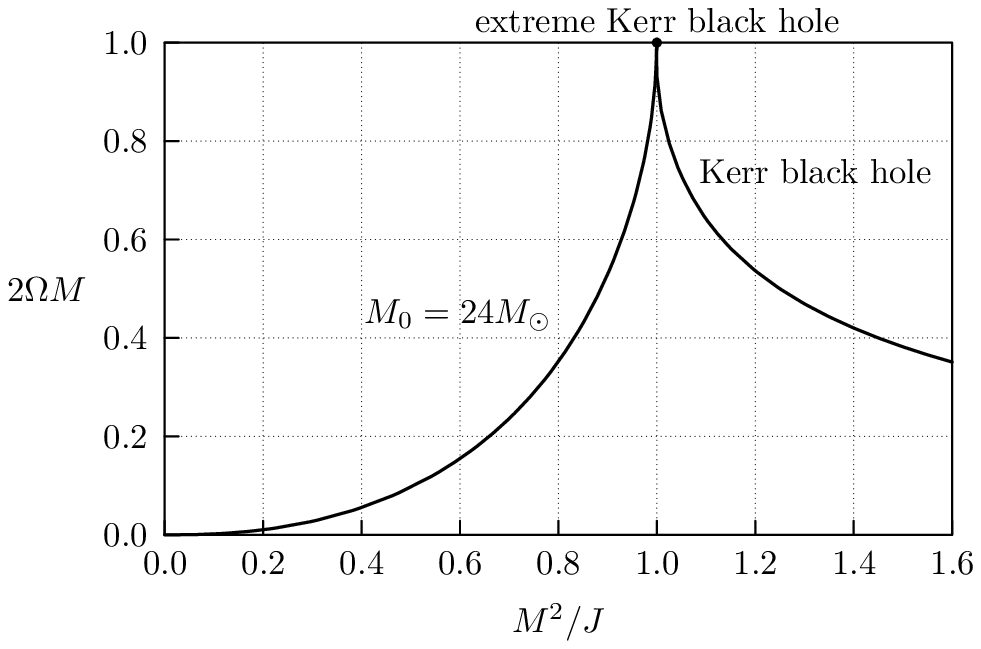}
	  \end{center}
	  \vspace{-1\baselineskip}
	  \caption{\label{Abb3}  
	  In the $r_1/r_2$-$Z_0/(Z_0+1)$ plane the mass-shedding sequence 
	  for rings governed by Chadrasekhar's equation of state is shown (cf. Fig.~\ref{Abb1}).
	  Moreover the specific sequence of such rings with fixed $M_0=24M_\odot$ is given.
	  In the lower panel this sequence is met by the sequence of Kerr black holes.}
	\end{figure}

	\begin{figure} 
	  \begin{center}
	    \includegraphics[scale=0.8]{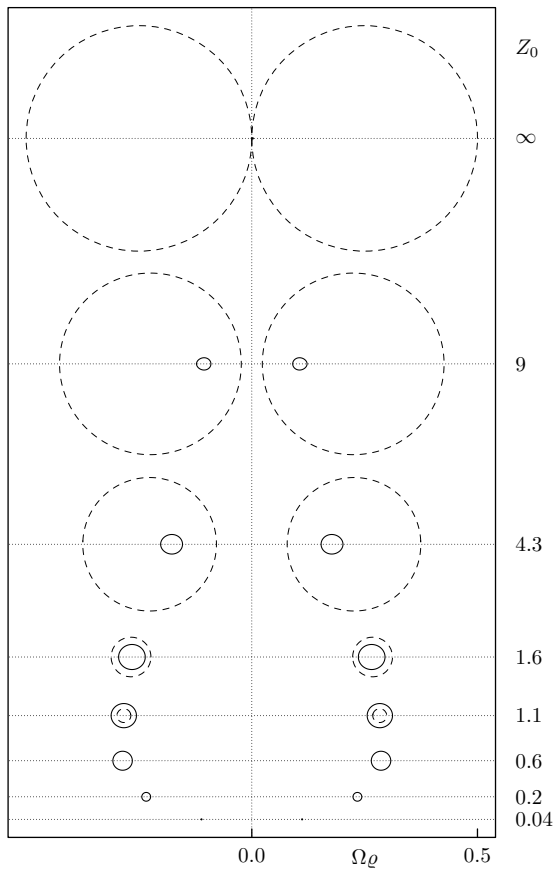}
	  \end{center}
	  \vspace{-1\baselineskip}
	  \caption{\label{Abb4}
	  		Meridional cross sections (solid) and ergospheres (dashed) for the
			ring sequence $M=24M_\odot$ governed by Chadra\-sekhar's equation of state. 
			The normalized $\zeta$-coordinate 
		   	$\Omega\zeta$ is plotted against the normalized $\varrho$-coordinate $\Omega\varrho$.
		   	In the limit $Z_0\to 0$ as well as in the extreme Kerr black hole 
			limit ($Z_0\to\infty$) the ring shrinks down to the normalized coordinate origin.}
	\end{figure}

	As shown by \citet{Meinel04}, the extreme Kerr solution is the only black hole limit 
	of rotating perfect fluid bodies in equilibrium. It is characterised by an infinite redshift $Z_0$. 
	Indeed, if a sequence of fluid configurations possesses no upper bound for $Z_0$, then  
	this sequence necessarily admits the transition to the extreme Kerr solution, see \citet{Meinel05}.
	
	However, spheroidal configurations do not seem to exhibit this limit. In particular, 
	for spherically symmetric static bodies $Z_0<2$ always holds, which is a consequence of the well-known 
	Buchdahl-limit. Numerical investigations have shown that $Z_0<7.378$ for the class of 
	uniformly rotating homogeneous spheroidal bodies \citep{SA03}. 
	The corresponding critical configuration with this maximal 
	redshift rotates at the mass-shedding limit and possesses infinite central pressure. 
	Although the mass-shedding limit is characteristic of maximal redshift configurations,
	infinite central pressure is not a typical feature
	(in particular, neither for Chandra\-sekhar's nor for polytropic equations of state with 
	$N>0$ does maximal redshift coincide with infinite central pressure).
	It is interesting to note that the value of the maximal redshift decreases with increasing polytropic 
	index $N$ (therefore $7.378$ is an upper bound). 
	The maximal redshift for Chandra\-sekhar's equation of state ($Z_0\approx 0.217$) is quite close 
	to the maximal redshift for the polytropic configurations with $N=3/2$ ($Z_0\approx 0.279$),
	cf. the discussion at the end of \ref{ChandEOS}.
 
	In contrast, there is no upper bound for the redshift of discs of dust \citep{NM95} and fluid rings \citep{AKM03c}.
	As mentioned at the end of section \ref{Metric}, the corresponding limiting procedure can be performed in two 
	different ways: 
	\begin{enumerate}
		\item
			If we consider some point in space outside the coordinate origin and hold the corresponding
			coordinates $(\rho,\zeta)$ fixed during the limiting process, then we notice that the equilibrium
			configuration shrinks down to the coordinate origin and the spacetime assumes the exterior
			geometry of the extreme Kerr solution in Weyl coordinates.
		\item
			If, on the contrary, we rescale the spatial coordinates such that $(\rho/r_2)$ and $(\zeta/r_2)$
			remain finite (for more details see \citet{Meinel02}, equations (47), (54) therein),
			we obtain a solution of Einstein's field equations that still describes an axisymmetric and stationary
			gravitational source in vacuum, but is no longer asymptotically flat. This interior, non-asymptotically flat
			solution is not unique. For each equation of state and a sufficiently large coordinate radius ratio 
			(see figure \ref{Abb1})
			we get a distinct interior space time. Moreover, in general a space time of this kind does not
			necessarily contain a gravitational source of toroidal topology. 
			As an important example of this, consider the infinite redshift limit
			of the analytically known solution corresponding to a rigidly rotating disc of dust \citep{NM95}.
			In this limit the gravitational source remains a disc whence the corresponding topology is different 
			from a ring topology.
	\end{enumerate}
	As pointed out in \citet{Meinel02}, the above two limits of space
	time geometries are disconnected from one another (the connection would be an ``infinitely extended throat'').
	In this way it is possible to obtain very different non-asymptotically flat interior limits and always
	the same exterior, extreme Kerr Black Hole limit.
	
	In this section we want to consider a specific sequence of configurations 
	governed by Chandrasekhar's equation of state, which allows the parametric transition
	to the extreme Kerr black hole. Along this sequence we hold the rest mass $M_0=24M_\odot$ fixed.

	From figure \ref{Abb3} we see that configurations with extremely small redshift $Z_0$ approach the 
	limit of infinitely thin rings in the Newtonian limit. That underlines once more the sophisticated 
	character of this limit: any sequence of configurations with bounded mass cannot terminate at some 
	$Z_0>0$ since all the points $(r_1/r_2=1, Z_0>0)$ correspond to configurations with infinite 
	masses (see figure \ref{Abb2}). 
	On the other hand, a sequence of finite mass configurations, $M_0>0$, cannot, of course, terminate in 
	a Newtonian limit. The only remaining possiblity is for the curve in diagram \ref{Abb3} 
	to move ``somewhere'' 
	between these limiting curves. Thus the limit of infinitely thin rings is more complex than the 
	figures \ref{Abb1} and \ref{Abb3} reveal. 

	As the redshift of the configurations in question increases, the ratio $r_1/r_2$ decreases to some minimal value
	$r_1/r_2\approx 0.7$. At this point the sequence approaches the extreme Kerr black hole limit in a way
	that is very similar to the example with $r_1/r_2=0.7$ discussed in \citet{AKM03c}. 
	This is confirmed by figure \ref{Abb4} in which meridional cross sections for selected 
	configurations of this sequence are shown. Note that ergospheres appear above a certain redshift,
	which again is in agreement with the results obtained in \citet{AKM03c}. 

{\par\bigskip\small Acknowledgements. We would like to thank R. Meinel and D. Petroff for many valuable discussions and 
helpful advice. This research was funded in part by the Deutsche Forschungsgemeinschaft (SFB/TR 7 - B1).\par}

\bibliographystyle{mn2e}
\bibliography{B1} 

\label{lastpage} 
 
\end{document}